\documentclass[9pt,twocolumn,twoside]{PNAS_template}
\articletype{CLASSIFICATION}%%%% Article topic/classification

\templatetype{PNAS_research_article} % Choose template
\usepackage{braket}
\usepackage{comment}
\usepackage{ulem}
\usepackage{xcolor}
\usepackage{ulem}

\begin{document}

\title{A universal low-temperature fluctuation of unconventional superconductivity revealed: Smoking gun favors Galilean bosonic superfluidity}

% Use letters for affiliations, numbers to show equal authorship (if applicable) and to indicate the corresponding author
\author[a,1]{Jie Wang}
\author[a,b,1]{Ruoshi Jiang}
\author[a,1]{Anthony Hegg}
\author[a]{Jinning Hou}
\author[a]{Tao Zeng}
\author[c]{Yucel Yildirim}
\author[a,d,e,2]{Wei Ku}

\affil[a]{School of Physics and Astronomy \& Tsung-Dao Lee Institute, Shanghai Jiao Tong University, Shanghai 200240, China}
\affil[b]{Department of Materials Science and Metallurgy, University of Cambridge, Cambridge CB3 0FS, United Kingdom}
\affil[c]{Department of Energy Systems Engineering, Bilgi University, Istanbul 34060, Turkey}
\affil[d]{Key Laboratory of Artificial Structures and Quantum Control (Ministry of Education), Shanghai 200240, China} 
%\affil[e]{Shanghai Branch, Hefei National Laboratory, Shanghai 201315, People's Republic of China}
\affil[e]{Shanghai Branch, Hefei National Laboratory, Shanghai 201315, China}

% Please give the surname of the lead author for the running footer
\leadauthor{Wang}

% Please add a significance statement to explain the relevance of your work
\significancestatement{Unconventional superconductivity has remained an unsolved mystery since its inception several decades ago.
The low-temperature superfluid density in these materials is the key property that illuminates the correct quantum theory while eliminating others.
Based on data from almost 20 families of bulk $3D$ superconductors, we identify a universal $T^3$ reduction of the superfluid density that rules out all previously proposed theories of superflow.
We then develop a Galilean invariant theory for hydrodynamic superfluidity that generically reproduces this $T^3$ depletion.
This result leaves charged superfluidity of emergent bosons as the most natural (and perhaps the only known) explanation for superconductivity in these materials.}

% Please include corresponding author, author contribution and author declaration information
\authorcontributions{
W. Ku designed the research; 
J. Wang, R. Jiang, A. Hegg, and W. Ku developed the theory; 
J. Hou, T. Zeng, and Y. Yildirim contributed to the theoretical analysis; 
R. Jiang analyzed the experimental data; 
J. Wang, R. Jiang, A. Hegg, and W. Ku wrote the paper.
}
\authordeclaration{The authors declare no competing interest.}
\equalauthors{\textsuperscript{1}J.W., R.J., and A.H. contributed equally to this work.}
\correspondingauthor{\textsuperscript{2}To whom correspondence should be addressed. E-mail: weiku@sjtu.edu.cn}

% At least three keywords are required at submission. Please provide three to five keywords, separated by the pipe symbol.
\keywords{unconventional superconductor $|$ superfluid $|$ many-body physics}

\begin{abstract}
Low-temperature thermal fluctuations offer an essential window in characterizing the true nature of a quantum state of matter, a quintessential example being Fermi liquid theory.
In the long-standing pursuit toward a fundamental understanding of unconventional superconductivity, the most essential low-temperature thermal fluctuation of superfluidity has unfortunately been greatly overlooked.
Here, we examine the leading thermal fluctuation of the superfluid density across numerous families ranging from relatively conventional to highly unconventional superconductors (MgB$_2$, bismuthates, doped buckyballs, heavy fermions, UTe$_2$, doped SrTiO$_3$, Chevrel clusters, intermetallics, $\kappa$-organics, transition metal dichalcogenides, ruthenates, iron-pnictides, cuprates, and kagome metals).
Amazingly, in all of them an unprecedented universal $T^3$ depletion materializes in the low-temperature superfluid density, even in the believed-to-be-conventional MgB$_2$.
This reveals a new quantum superfluid state of matter and requires a necessary change of paradigm in describing these modern superconductors.
We demonstrate that such an unorthodox yet generic behavior can be described by a Galilean invariant theory of hydrodynamic superfluidity hosting a long-lived `true condensate'.
\end{abstract}

\dates{This manuscript was compiled on \today}
\doi{\url{www.pnas.org/cgi/doi/10.1073/pnas.XXXXXXXXXX}}

\maketitle
\thispagestyle{firststyle}
\ifthenelse{\boolean{shortarticle}}{\ifthenelse{\boolean{singlecolumn}}{\abscontentformatted}{\abscontent}}{}

\Firstpage

%The \Firstpage command is used to format the first page text column size. The same size will be maintained for subsequent paragraph until the \Endparasplit or \Parasplit command is encountered.

Superconductivity has remained one of the most important topics in condensed matter physics since its discovery long ago~\cite{Kapitza,AllenMisener,Bogoliubov,LandauSF2Short}. Although traditional `conventional' superconductors seem to be well-described by the Bardeen-Cooper-Schrieffer (BCS) theory~\cite{BCS} as Fermi liquids with spontaneously broken U(1) symmetry, in the modern era a vast and ever-growing array of `unconventional' superconductors have been discovered that display behaviors beyond the applicability of the standard lore. A satisfactory understanding of the coherent quantum state of matter behind this purely quantum phenomenon therefore remains an open question without consensus.

The most essential and experimentally accessible properties of such a \textit{quantum} state are its thermal fluctuations at low temperature.
Given that qualitatively different fluctuations arise from distinct quantum states of matter, they offer a `smoking gun' signature that quickly falsifies any theories with incorrect ground states.
On the other hand, around a ground state with otherwise correct macroscopic properties, consistency of the characteristic fluctuations validates the effectiveness of a theory.
This greatly contrasts with the obfuscating effects of universality in the vicinity of a phase transition, where many qualitatively different models can exhibit identical behavior.
A good analogy to such a state is the 3D Fermi liquid, whose quantum nature is characterized by $T^2$ resistivity,  $T^2$ one-body scattering rate, $T^2$ Hall angle, $T$-linear specific heat, $1/T$ thermal conductivity, $T$-independent magnetic susceptibility, and $T$-independent Hall coefficient~\cite{AshcroftMermin}.
These universal fluctuations provide unambiguous criteria to determine whether a material is described by a particular known quantum state of matter (e.g. Fermi liquid) or not (i.e. non-Fermi liquid).

In terms of superconductivity, the most direct characterization is naturally through \textit{low-temperature} thermal fluctuations of the superfluid density (or \Parasplit correspondingly, the penetration depth). Particularly, through continuous improvement in the experimental setup, an unprecedented resolution for the penetration depth at low temperature and over a wide range of carrier density has been reached [e.g., in Ref.~\cite{hardyUDCuprate}]. Unfortunately, unlike studies on several other indirect quantities like resistivity, magnetic susceptibility and specific heat [e.g., Refs.~\cite{lohneysen1994non,taillefer,hardyspecheat}], a careful analysis of the low-temperature superfluid density has been largely overlooked by the community in favor of the multi-gap fitting over the entire superconducting temperature range~\cite{2MgB2_2,4LiFeAs,8PrPt4Ce14,6BaFe2As2,7LaNiGa2,18Mo5PB2,3CsV3Sb5}. As such, in the absence of a careful characterization of the superfluid properties in the low-temperature limit, there is simply not enough guidance to reliably identify a compatible quantum state of matter, aside from ruling out some obvious scenarios (such as single-gap BCS). 

Here, we carefully perform this essential analysis based on the available low-temperature penetration-depth data from a large number of superconductor families, ranging from the conventional to the highly unconventional, including MgB$_2$, bismuthates, doped buckyballs, heavy fermions, UTe$_2$, doped SrTiO$_3$, Chevrel clusters, intermetallics, organic superconductors, transition metal dichalcogenides, ruthenates, iron pnictides, cuprates, and kagome metals.
Surprisingly, in all of these compounds (both fully-gapped and nodal), the low-temperature superfluid density displays a universal $T^3$ depletion.
Mysteriously, this is beyond that of the standard BCS state and more pronounced than that of the traditional bosonic superfluid state.
This unprecedented universal behavior therefore precludes the corresponding quantum states in the existing theories and reveals a new quantum superfluid state of matter requiring a change of paradigm in describing modern superconductors.

To this end, we develop a strictly Galilean invariant quantum theory of superfluidity beyond the current lore.
Through employment of \textit{long-lived} fully-dressed eigen-particles of general interacting bosonic systems, we introduce a rigorous `true condensate' and associated with it an exact microscopic definition of supercurrent.
Specifically, via the rigorous analytical structure of the current fluctuation, we identify beyond the standard sound velocity the essential inertial mass velocity responsible for the net current.
This addition not only enables proper Galilean frame transformation for the elementary excitations, but also straightforwardly establishes an \textit{equivalence} between the hydrodynamic superfluid density and the true condensate density for an arbitrary quantum state.
The thermal depletion of the superfluid density in such a proper microscopic superfluid theory thus follows the $T^3$ thermal fluctuation of the three-dimensional true condensate.
To the best of our knowledge, this strictly Galilean invariant theory of bosonic superfluidity is the only remaining theory capable of describing the quantum nature of the unconventional superconductivity in these materials.

\section*{Results}\label{results}

\subsection*{Unorthodox universal low-temperature fluctuation}\label{fitlowT}

\begin{figure}[t!]
    \centering
    \includegraphics[width=8.7cm]{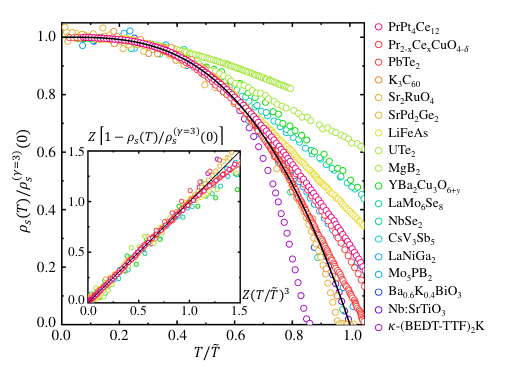}
    \caption{
\textbf{Universal $T^3$ thermal depletion of superfluid density}
Experimentally observed superfluid density is analyzed via data-collapse to $\rho_\mathrm{s}(T)/\rho_\mathrm{s}^{(\gamma=3)}(0)$ vs rescaled temperature $T/ \tilde{T}$ and to $1-\rho_\mathrm{s}(T)/\rho_\mathrm{s}^{(\gamma=3)}(0)$ vs $(T/ \tilde{T})^3$ (inset) using extrapolated zero-temperature superfluid density $\rho_\mathrm{s}^{(\gamma=3)}(0)$ and material-dependent rescaling factor $\tilde{T}$. For easier visualization, the horizontal and vertical axes of the inset are rescaled by a material-dependent $Z$ factor (c.f. Table~\ref{zoom} in Materials and Methods) such that the $T^3$ behavior dominates below the value $1.0$ in the axes. A clear cubic $T^3$-depletion is found at low temperature for all displayed materials.
Here, $\mathrm{x}$ = 0.131 in $\mathrm{Pr_{2-x}Ce_xCuO_{4-\delta}}$ and $\mathrm{y}$ = 0.419 in $\rm{YBa_2Cu_3O_{6+y}}$, while the carrier density in $\mathrm{Nb}$:$\mathrm{SrTiO}_3$ is 0.33 $\times$ 10$^{20}$ cm$^{-3}$ and $\mathrm{K}$ denotes $\mathrm{Cu[N(CN)_2]Br}$ in the organic superconductor $\kappa$-$\mathrm{(BEDT-TTF)_2K}$.}\label{expfit}
\end{figure}

Figure~\ref{expfit} reveals an incredible \textit{universal} low-temperature fluctuation in nearly all modern superconductor families.
The figure displays the observed bulk 3D low-temperature superfluid density $\rho_\mathrm{s}$ for many materials, found earlier~\cite{17BKBO,12K3C60,9PCCO,2MgB2_2,5NbSe2,hardyUDCuprate,4LiFeAs,8PrPt4Ce14,14SrPd2Ge2,6BaFe2As2,15Organic} and more recently~\cite{7LaNiGa2,11PbTe2,18Mo5PB2,3CsV3Sb5, 19Sr2RuO4,20UTe2,21STO}, data-collapsed to a single function via $\rho_\mathrm{s}(T)/\rho_\mathrm{s}^{(\gamma=3)}(0)$ vs rescaled temperature $T/ \tilde{T}$ using extrapolated zero-temperature superfluid density $\rho_\mathrm{s}^{(\gamma=3)}(0)$.
(Details are available in Materials and Methods. Note that unlike the universal behavior around the thermal phase transition, low-temperature scaling does not have an obvious scale as a convenient unit, hence the material-dependent rescaling factor $\tilde{T}$.)
The inset displays the same data, now collapsed to $1-\rho_\mathrm{s}(T)/\rho_\mathrm{s}^{(\gamma=3)}(0)$ as a function of $(T/ \tilde{T})^3$ as an aid to the eye.
Amazingly, \textit{all} of them fall into a universal $T^3$ power-law form at low enough temperature.
Particularly, K$_3$C$_{60}$ and Nb-doped SrTiO$_3$ even
maintain a $T^3$ form over the entire superconducting phase!
In fact, the same universal $T^3$ form also applies to doped YBa$_2$Cu$_3$O$_7$ (c.f. SI Appendix, Section 1), \textit{independent} of doping level around the quantum critical point, where high-quality data ($10^{-4}$ cation disorder) is available~\cite{hardyUDCuprate}.
This unexpected \textit{quantum} universality spans virtually all of the superconductor families studied over the past several decades and certainly requires an equally general theory to account for it.

%Figure 1

\subsection*{Direct physical implications}\label{thyimp}

However, such a power-law fluctuation, as opposed to gap-protected exponential behavior, is unexpected for fully charge-gapped superconductors~\cite{Emery1995} such as MgB$_2$, B$_{0.6}$K$_{0.4}$BiO$_3$, LaMo$_6$Se$_8$, Nb-doped SrTiO$_3$ and K$_3$C$_{60}$.
Particularly, the 3D nature of the \textit{low-temperature} superconductivity in these materials renders inapplicable the gapless phase-fluctuation scenario~\cite{DoniachPhzFluc,EmeryKivelson}, whereas the Anderson-Higgs mechanism would give a fully gapped phase mode~\cite{anderson-higgs}.
Therefore, one has \textit{no choice} but to conclude the \textit{absence} of the Anderson-Higgs mechanism in these materials.

This absence of the Anderson-Higgs mechanism might seem alarming to some experts in superconductivity.
However, recall that this mechanism, like nearly all physical mechanisms, is established only under a series of assumptions or preconditions.
The observed power-law fluctuation therefore implies the invalidation of at least one of these assumptions or preconditions.
For example, in the presence of strong electronic correlations, especially with tightly bound bi-electrons of short coherence length in these strongly type-II superconductors, the low-energy dynamics would contain a \textit{constrained} Hilbert space~\cite{t-J_model} and non-perturbative emergent interactions \textit{indescribable} by photon exchange.
Such systems are thus \textit{beyond} the scheme of the Anderson-Higgs mechanism (c.f. SI Appendix, Section 2 for a prototypical example).

Even more puzzling is the same $T^3$-depletion displayed in Fig.~\ref{expfit} in gapless superconductors such as $\mathrm{YBa_2Cu_3O_{6+y}}$ (y = 0.419), $\mathrm{Pr_{2-x}Ce_xCuO_{4-\delta}}$(x = 0.131) and $\mathrm{Ba(Fe_{0.9}Co_{0.1})_2As_2}$.
Since thermal population of gapless Bogoliubov quasi-particles is inescapable, gapless BCS theories generally would result in $T$-linear (line node) or $T^2$ (point node) depletion of the superfluid density\footnote{Higher powers can only result from specially fine-tuned pairing combinations with specific applied field directions.}~\cite{AGR1991}, given that a Bogoliubov quasi-particle represents exciting a fermion from Cooper pairs in the superflow.
It is therefore yet again inconceivable how, for example, $p$-wave and/or $d$-wave BCS theories could ever account for such results.
The only remaining scenario is that, contrary to all existing pairing theories, in those gapless superconductors~\cite{Won2005,Tsuei1994, ARPESnodal, 15Organic_nodal} the low-energy fermions were \textit{never} part of the superflow (e.g. part of a Cooper pair) even before being excited!
%In turn, this unorthodox scenario now validates a previous proposal~\cite{prepairmodelweiku} in which unpaired fermions, instead of joining the superflow, form a `condensate polaron' dressed by a bosonic condensate via coherent scattering.
In turn, this unorthodox behavior validates a previous proposal~\cite{prepairmodelweiku} for a second type of superconducting quasi-particle gap, in which the Bogoliubov-like excitation removes a bare fermion \textit{not} from a Cooper pair, but instead from a `super-polaron', a fermion coherently dressed by a bosonic superfluid.

However, for a 3D-coherent bosonic superflow with gapless phase mode, the traditional hydrodynamic theory~\cite{Landau_vol9} exhibits $T^4$ depletion instead, close but not identical to the above $T^3$ depletion.
%On the other hand
Moreover, the $T^3$ fluctuation of 2D superfluidity is inapplicable here, as these materials all demonstrate 3D coherent superconductivity at temperatures lower than the coherent energy scale of the third dimension.
It is therefore necessary to \textit{meticulously} re-visit the thermal depletion of the superfluid density.

% %\subsection{Rigorous bosonic superfluid theory and generic $T^3$-depletion}
\subsection*{Rigorous structure of superfluid and generic $T^3$ depletion}

To this end, we develop a Galilean invariant, number-conserving superfluid theory for quantum interacting bosonic systems with translational symmetry via employment of fully-dressed long-lived `eigen-particles'~\cite{hegg1,Steven_White,KannoI}, $\tilde{a}^{\dag}_{\mathbf{k}}$, that carry the same inertial mass $m$ and momentum $\hbar\textbf{k}$ of the bare particles (c.f. Materials and Methods).

Consider a general Galilean invariant $N$-particle interacting bosonic system as a low-energy \textit{effective} description of the systems in Fig.~\ref{expfit},
\begin{align}
\label{bare_H}
H = \sum_{\mathbf{k}_1} \epsilon_{\mathbf{k}_1} a^{\dag}_{\mathbf{k}_1} a_{\mathbf{k}_1} + \frac{1}{2}\sum_{\mathbf{q}~\mathbf{k}_1\mathbf{k}_2 } V_\mathbf{q} a^{\dag}_{\mathbf{k}_1+\mathbf{q}} a^{\dag}_{\mathbf{k}_2-\mathbf{q}} a_{\mathbf{k}_2} a_{\mathbf{k}_1},
\end{align}
represented by second-quantized creation operators $a^{\dag}_{\mathbf{k}}$ for itinerant carriers of momentum $\mathbf{k}$\footnote{Throughout this manuscript, discrete momenta are employed to facilitate careful counting of the Hilbert space and states.} having kinetic energy $\epsilon_{\mathbf{k}}=\frac{\hbar^2}{2m}\mathbf{k}\cdot\mathbf{k}$. Importantly, as constrained by the above experimental observation, the effective interaction $V_\mathbf{q}$ is limited to be distinct from the bare Coulomb interaction (i.e. incompatible with the simple exchange of photons), such that the Anderson-Higgs mechanism is inapplicable to such \textit{non-perturbatively} emergent tightly-bound bosons with strict \textit{number conservation} in their dynamics.
% %As to be elaborated below, an essential difficulty in the existing quantum descriptions~\cite{Landau_vol9,HuangYang,LeeHuangYang} is to identify a \textit{long-lived} curl-less portion of the current in a \textit{particle-conserving} system as the superfluid component.
As will be elaborated below, an essential difficulty in existing quantum descriptions~\cite{Landau_vol9,HuangYang,LeeHuangYang} is to identify a \textit{long-lived} supercurrent component in a \textit{particle-conserving} system.
To this end, we resort to the diagonal-representation of $H$~\cite{hegg1,KannoI},
\begin{align}
\label{Hamiltonian}
H_{\text{diag}}&[\{ \tilde{a}^{\dag}_{\mathbf{k}}\}] \equiv H[\{a^{\dag}_{\mathbf{k}}\}]\\
\nonumber
=& \sum_{\mathbf{k}_1} \epsilon_{\mathbf{k}_1} \tilde{a}^{\dag}_{\mathbf{k}_1} \tilde{a}_{\mathbf{k}_1} + \frac{1}{2!}\sum_{\mathbf{k}_1 \mathbf{k}_2} \epsilon_{\mathbf{k}_1 \mathbf{k}_2} \tilde{a}^{\dag}_{\mathbf{k}_1} \tilde{a}^{\dag}_{\mathbf{k}_2} \tilde{a}_{\mathbf{k}_2} \tilde{a}_{\mathbf{k}_1} \\
\nonumber
+& \cdots +  \frac{1}{N!} \sum_{\mathbf{k}_1 \cdots \mathbf{k}_N} \epsilon_{\mathbf{k}_1 \cdots \mathbf{k}_N} \tilde{a}^{\dag}_{\mathbf{k}_1} \cdots \tilde{a}^{\dag}_{\mathbf{k}_N} \tilde{a}_{\mathbf{k}_N} \cdots \tilde{a}_{\mathbf{k}_1},
\end{align}
using the long-lived fully-dressed bosonic `eigen-particles',
$\tilde{a}^{\dag}_{\mathbf{k}} \equiv U^{\dag} a^{\dag}_{\mathbf{k}} U$,
transformed from the `bare' particle creation operators $a^{\dag}_{\mathbf{k}}$ via a unitary operator $U$.
% %The canonical commutation, $[\tilde{a}_{\mathbf{k}},\tilde{a}^{\dag}_{\mathbf{p}}] = \delta_{\mathbf{k}\mathbf{p}}$, testifies to eigen-particles' bosonic nature.

The eigen-particles allow unprecedented \textit{direct} access to properties of long time-scale, since $\tilde{N}_\mathbf{k}(t)\equiv\tilde{a}^{\dag}_{\mathbf{k}}(t)\tilde{a}_{\mathbf{k}}(t)$ in the Heisenberg picture is a constant of motion, $\tilde{N}_{\mathbf{k}}(t)=\tilde{N}_{\mathbf{k}}$ given $[\tilde{N}_{\mathbf{k}},H_\mathrm{diag}]=0$.
By fully absorbing interaction-induced quantum fluctuations (of finite time scale) into their internal structure, eigen-particles are preserved during time evolution, as evident from the diagonal structure of $H_\textrm{diag}$.

Since $U$ is tasked to bring only the beyond-one-body terms in $H$ into a diagonal form, the coefficients $\epsilon_{\mathbf{k}}$ of the \textit{already diagonal} one-body terms in \eqref{bare_H} are preserved in \eqref{Hamiltonian}, independent of its precise form.
Furthermore, Galilean invariance of the system introduces a symmetry property, $\epsilon_{\mathbf{k}_1+\mathbf{k}_0,\mathbf{k}_2+\mathbf{k}_0,\cdots}=\epsilon_{\mathbf{k}_1,\mathbf{k}_2, \cdots}$,
that ensures \textit{kinematic} momentum conservation throughout many-body (c.f. Materials and Methods).

\begin{SCfigure*}[\sidecaptionrelwidth][t!]
        \centering
     \includegraphics[width=11.4cm]{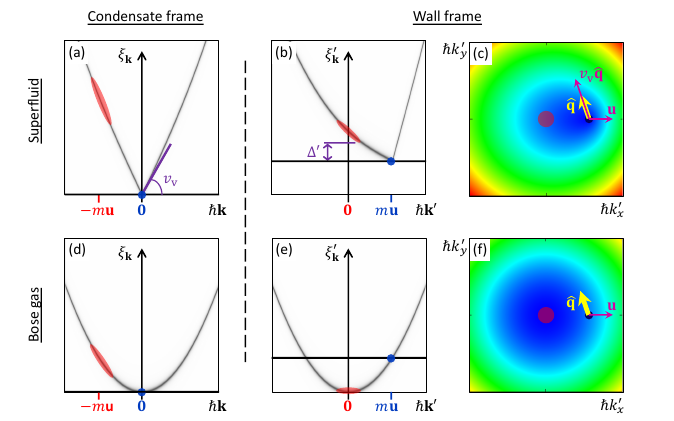}
	\caption{
 \textbf{Illustration of superfluid and its stiffness} in equilibrium with a moving wall. (a) Momentum $\hbar\mathbf{k}$ dependent eigen-particle removal energy $\xi_\mathbf{k}$, shown in the rest frame of the true condensate (in blue), contains an extra contribution of stiffness-enabled vibrational energy $v_\mathrm{v}\hbar|\mathbf{k}|$ in comparison with the Bose gas in (d).
 (b) In the wall frame, the condensate moves with a steady velocity $\mathbf{u}$, while the uncondensed eigen-particles (in red) in equilibrium with the wall stay stationary, similar to the case of Bose gas in (e).
 Notice that the stiffness-enabled vibrational energy in a superfluid results in a finite energy $\Delta^\prime$ (and momentum) cost for condensed particles to slow down and reach equilibrium with the wall and thereby prohibits their dissipation.
 (c) Contour plot of $\xi_\mathbf{k}$ in the wall frame reveals an extra longitudinal-only component $v_\mathrm{v}\hat{\mathbf{q}}\big|_{\mathbf{q=k-k}_0}$ in the eigen-particle velocity $\mathbf{\nu}_\mathbf{k}=\frac{1}{\hbar} \nabla_{\mathbf{k}}\xi_\mathbf{k}$ around the true condensate of momentum $\mathbf{k}_0$, in comparison with the gas in (f).
 This indicates that the true condensate, and only the true condensate is fully protected by a minimum velocity in \textit{every} direction to survive the thermal equilibration process as the hydrodynamic superfluid component.
 }
	\label{fig_supercurrent}
\end{SCfigure*}

Evidently from \eqref{Hamiltonian}, all naturally thermalized eigenstates of the system are simply \textit{product} states of $N$ eigen-particles,
\begin{align}
\ket{\{N_{\mathbf{k}} \}} &\equiv \prod_{\mathbf{k}} \frac{( \tilde{a}^{\dag}_{\mathbf{k}})^{N_{\mathbf{k}}}}{\sqrt{N_{\mathbf{k}}!}} \ket{0}\Big|_{\sum_{\mathbf{k}}N_{\mathbf{k}}=N}~.
\label{eigenstates}
\end{align}
For standard Galilean invariant, kinetic energy-dominated superfluid systems, the ground state is therefore a `true Bose-Einstein condensate'
$\ket{\Omega} \equiv \frac{1}{\sqrt{N!}}( \tilde{a}^{\dag}_{\mathbf{k}_0} )^N \ket{0}$,
containing all $N$ eigen-particles residing at momentum $\mathbf{k}_0=\mathbf{0}$ with the lowest kinetic energy [the blue dot in Fig.~\ref{fig_supercurrent}(a)].
Similarly, the elementary excited states are those with eigen-particles excited from the true condensate [for example to the red region in Fig.~\ref{fig_supercurrent}(a)], such as $\ket{E_\mathbf{k}} = \frac{1}{\sqrt{(N-1)!}}\tilde{a}^{\dag}_\mathbf{k}( \tilde{a}^{\dag}_{\mathbf{k}_0} )^{N-1} \ket{0}$.

The above unconventional definition of `true condensate' makes perfect physical sense, since $\tilde{N}_0(t)=\tilde{N}_0\equiv\tilde{a}^{\dag}_{\mathbf{k}_0}\tilde{a}_{\mathbf{k}_0}$ is a constant of motion, consistent with condensation being a long-lived phenomenon.
In contrast, the standard `bare' condensate, $a^{\dag}_{\mathbf{k}_0}(t)a_{\mathbf{k}_0}(t)$, always fluctuates on a much shorter time scale ($[a^{\dag}_{\mathbf{k}_0}a_{\mathbf{k}_0},H]\neq0$) and is therefore not directly applicable to any long-lived property of the system.

As well-established in the literature~\cite{nozieres1999}, the superfluid's resilience against dissipation is phenomenologically associated with the emergence of stiffness that restricts the allowed excitations to only long-lived elementary ones of \textit{positive} energy in \textit{all} inertial frames close to the superfluid frame.
This corresponds to an extra interaction-induced `vibrational energy' contribution $v_\mathrm{v}\hbar|\mathbf{k}|$~\cite{nozieres1999} to the eigen-particle removal energy (c.f. Materials and Methods), $\xi_\mathbf{k}\equiv\braket{\frac{1}{\tilde{N}_\mathbf{k}}\tilde{a}^\dagger_\mathbf{k}\tilde{\epsilon}_{\mathbf{k}}\tilde{a}_\mathbf{k}}$, shown in Fig.~\ref{fig_supercurrent}(a).
Here, the \textit{diagonal} $\tilde{\epsilon}_{\mathbf{k}}$ operator is defined via $\tilde{\epsilon}_{\mathbf{k}}\tilde{a}^\dagger_\mathbf{k}\equiv[H,\tilde{a}^\dagger_\mathbf{k}]$,
\begin{align}
\tilde{\epsilon}_{\mathbf{k}}
%&= \epsilon_{\mathbf{k}} + \sum_{\mathbf{k}_1}\epsilon_{\mathbf{k}\mathbf{k}_1} \tilde{a}^{\dag}_{\mathbf{k}_1} \tilde{a}_{\mathbf{k}_1}+ \frac{1}{2!}\sum_{\mathbf{k}_1\mathbf{k}_2}\epsilon_{\mathbf{k}\mathbf{k}_1\mathbf{k}_2} \tilde{a}^{\dag}_{\mathbf{k}_1} \tilde{a}^{\dag}_{\mathbf{k}_2} \tilde{a}_{\mathbf{k}_2} \tilde{a}_{\mathbf{k}_1} + \cdots \nonumber\\
%&\equiv
=\epsilon_{\mathbf{k}}+\tilde{\Sigma}_\mathbf{k}=\frac{\hbar^2}{2m}\mathbf{k}\cdot\mathbf{k}+\tilde{\Sigma}_\mathbf{k},
\label{epsilon_tilde}
\end{align}
that includes the \textit{Galilean invariant} kinetic energy of inertial mass $\frac{\hbar^2}{2m}\mathbf{k}\cdot\mathbf{k}$ and the interaction-induced self-energy $\tilde{\Sigma}_\mathbf{k}$ that encapsulates the vibrational energy.

%[bare mass + lack of scattering (constant of motion) + energy protection --> $rho_s$ directly?]

The profound implications of the emergent stiffness on transport properties of long space-time scales are naturally described by the current fluctuation of eigen-particles.
Defined as the time derivative of the eigen-particles' spatial distribution (c.f. Materials and Methods), the dressed current fluctuation,
\begin{align}
\tilde{\mathbf{J}}_\mathbf{q} = \frac{1}{\mathcal{V}}\sum_{\mathbf{k}} \tilde{a}^{\dag}_{\mathbf{k+q}} \tilde{\mathbf{v}}_{\mathbf{k+\frac{q}{2}}} \tilde{a}_{\mathbf{k}}~,
\label{dressed_current}
\end{align}
of wavenumber $\mathbf{q}$ strictly satisfies the continuity equation at length scales $2\pi/\mathbf{q}$ longer than the size of eigen-particles.
Here, the diagonal velocity operator of the eigen-particles,
\begin{align}
\tilde{\mathbf{v}}_\mathbf{k}=\frac{1}{\hbar} \nabla_{\mathbf{k}} \tilde{\epsilon}_{\mathbf{k}}
%=\frac{1}{\hbar} \nabla_{\mathbf{k}}\big( \epsilon_{\mathbf{k}} + \tilde{\Sigma}_{\mathbf{k}}\big)
=\frac{\hbar\mathbf{k}}{m} + \frac{1}{\hbar} \nabla_{\mathbf{k}} \tilde{\Sigma}_{\mathbf{k}}~,
\label{full_velocity}
\end{align}
gives the eigen-particle velocity, $\mathbf{\nu}_\mathbf{k}\equiv\braket{\frac{1}{\tilde{N}_\mathbf{k}}\tilde{a}^\dagger_\mathbf{k}\tilde{\mathbf{v}}_{\mathbf{k}}\tilde{a}_\mathbf{k}}=\frac{1}{\hbar} \nabla_{\mathbf{k}}\xi_\mathbf{k}$.
As shown in Fig.~\ref{fig_supercurrent}(c), for pure states hosting true condensate of an arbitrary momentum $\hbar\mathbf{k}_0$, $\mathbf{\nu}_\mathbf{k}$ contains an extra isotropic vibration velocity $\mathbf{v}_\mathrm{v}(\mathbf{k})\equiv v_\mathrm{v}\hat{\mathbf{q}}\big|_{\mathbf{q=k-k}_0}$, in addition to the Galilean invariant inertial mass velocity $\frac{\hbar\mathbf{k}}{m}$ already present in the Bose gas in Fig.~\ref{fig_supercurrent}(f).

Notice that Galilean invariance of \eqref{dressed_current} is maintained by the kinetic contribution of inertial mass in \eqref{full_velocity}.
In the condensate frame, such contribution is sub-leading to the vibrational contribution and absent in existing sound-based theories~\cite{Landau_vol9,LeeHuangYang}.
However, it is the one that describes the essential inertial mass flow (such as the supercurrent).
This is most obvious in the net steady current, $\tilde{\mathbf{J}}$, since vibrations always have finite wavelength and thus do not contribute at $\mathbf{q}=\mathbf{0}$.
Indeed, the above-mentioned Galilean symmetry of $\epsilon_{\mathbf{k}_1\mathbf{k}_2, \cdots}$ ensures that the net current of the system,
\begin{align}
\tilde{\mathbf{J}} &=\sum_\mathbf{k}\frac{1}{\mathcal{V}} \tilde{a}^{\dag}_{\mathbf{k}} \tilde{\mathbf{v}}_{\mathbf{k}} \tilde{a}_{\mathbf{k}}
=\frac{1}{\mathcal{V}}\sum_{\mathbf{k}} \frac{\hbar\mathbf{k}}{m}\tilde{a}_{\mathbf{k}}^{\dag}\tilde{a}_{\mathbf{k}},
%=\frac{1}{m\mathcal{V}}\tilde{\mathbf{P}},
\label{J_total}
\end{align}
contains only the \textit{inertial mass flow}, as expected from the conserved kinematic momentum, $\tilde{\mathbf{J}}=\frac{1}{m\mathcal{V}}\tilde{\mathbf{P}}=\frac{1}{m\mathcal{V}}\mathbf{P}=\mathbf{J}$ .

To access the superfluid density defined in the \textit{hydrodynamic} ``two-fluid'' description~\cite{landau2013fluid}, one needs to split the current into a normal current component and a special supercurrent one capable of resisting dissipation against slow external drive throughout the \textit{entire} equilibration process.
Since such equilibration in a liquid typically involves internal vortices, it corresponds to low-velocity excitation of eigen-particles.
% From Fig.~\ref{fig_supercurrent}(b)(e), it is straightforward to see that for a given $\mathbf{q}$ generally all eigen-particles are excitable via $\tilde{\mathbf{J}}_\mathbf{q}$ with negligible energy (or speed $\Delta \xi/q$) along some directions.
From Fig.~\ref{fig_supercurrent}(b)(e), it is straightforward to see that for a given wavenumber $q$, generally all eigen-particles are excitable via $\tilde{\mathbf{J}}_\mathbf{q}$ with negligible energy $\Delta \xi$ (or speed $\Delta \xi/q$) along some directions.

Except, in the presence of the superfluid stiffness in Fig.~\ref{fig_supercurrent}(b), eigen-particles in the true condensate (in blue) are fully protected by a minimum velocity in \textit{every} direction, effectively disabling the current fluctuation,
\begin{align}
\tilde{\mathbf{J}}^\mathrm{s}_\mathbf{q} =\frac{1}{\mathcal{V}}(\tilde{a}^{\dag}_{\mathbf{k}_0+\mathbf{q}} \tilde{\mathbf{v}}_{\mathbf{k}_0+\frac{\mathbf{q}}{2}} \tilde{a}_{\mathbf{k}_0}+\tilde{a}^{\dag}_{\mathbf{k}_0} \tilde{\mathbf{v}}_{\mathbf{k}_0-\frac{\mathbf{q}}{2}} \tilde{a}_{\mathbf{k}_0-\mathbf{q}}),
\label{J_s_q}
\end{align}
of the true condensate in all low-velocity channels.
The dissipation-resilient net supercurrent,
\begin{align}
\label{J_s}
\tilde{\mathbf{J}}^\mathrm{s} &=\frac{1}{\mathcal{V}}\frac{\hbar\mathbf{k}_0}{m}\tilde{a}_{\mathbf{k}_0}^{\dag}\tilde{a}_{\mathbf{k}_0} ~~~~\textrm{(for $v_\mathrm{v}>0$)},
\end{align}
thus contains only the inertial mass flow of the eigen-particles in the true condensate.
Evidently from \eqref{J_s_q} and \eqref{J_s}, the superfluid density $\rho_\mathrm{s}$, defined as the inertial mass density of the superflow~\cite{landau2013fluid}, $\rho_\mathrm{s}\mathbf{v}_\mathrm{s}\equiv m\tilde{\mathbf{J}}^\mathrm{s}$, is proportional to the true condensation density, $\tilde{N}_0/\mathcal{V}$,
\begin{align}
\label{superfluid_density}
\rho_\mathrm{s} = m\tilde{N}_0 / \mathcal{V}~~~~(\mathrm{for}~v_\mathrm{v} > 0),
\end{align}
given $\mathbf{v}_\mathrm{s}=\frac{\hbar \mathbf{k}_0}{m}$ of condensed particles.
Note that this \textit{rigorous} quantum mechanical equivalence in \eqref{superfluid_density}
applies to \textit{arbitrary} quantum states (and thus any ensemble of them).

In contrast, without such protection, eigen-particles outside the true condensate would be \textit{repeatedly} excited throughout the equilibrium process, until eventually reaching an average velocity, $-\mathbf{u}$, of the wall [c.f. the red region in Fig.~\ref{fig_supercurrent}(a)] according to the standard boundary condition of dissipative liquids.
Parallel to the analysis above, their associated net current $-\tilde{N}_\mathrm{n}\mathbf{u}/\mathcal{V}$ with $\tilde{N}_\mathrm{n}\equiv \sum_\mathbf{k\neq\mathbf{0}} \tilde{a}_{\mathbf{k}}^\dagger \tilde{a}_\mathbf{k}$ must correspond to the normal current component, $\tilde{\mathbf{J}}^\mathrm{n}$, having a normal fluid density,
\begin{equation}
    \rho_\mathrm{n} = m \tilde{N}_\mathrm{n}/\mathcal{V}.
\label{normal_fluid_density}
\end{equation}
Notice that while defined independently, the above superfluid and normal fluid densities trivially satisfy the physical sum rule, $\rho_\mathrm{n} + \rho_\mathrm{s} = m \tilde{N} /\mathcal{V} = mN/\mathcal{V} \equiv \rho$.

This nice feature is in significant contrast to the standard definition of these densities via \textit{linear response} (LR) of the system~\citep{Baym_1968,HuangYang, LeeHuangYang}, in which only $\rho^\mathrm{(LR)}_\mathrm{n}$ can be directly extracted while $\rho^\mathrm{(LR)}_\mathrm{s}$ has to be inferred from the sum rule.
In fact, it should be apparent that our \textit{hydrodynamic} two-fluid definition~\cite{landau2013fluid} describes the steady state \textit{after} reaching an equilibrium with the wall.
%Such state is drastically distinct from the steady state \textit{without} the wall, as probed by the linear response.
Distinct from typical systems, for a superfluid such state is drastically different from that probed by the linear response, which is instead a steady state \textit{without} the wall. (c.f. Discussion)
The superfluid component in the former, $\rho_\mathrm{s}$, requires a complete protection throughout the \textit{entire} equilibrium process, while in the latter, $\rho^\mathrm{(LR)}_\mathrm{s}$, it only needs to retain its curl-less characteristics during the weak linear response process.
Therefore, our $\rho_\mathrm{s}$ must be smaller than $\rho^\mathrm{(LR)}_\mathrm{s}$ defined through linear response, but is exactly the one employed in the London equation~\citep{London_1935} for those experimental analyses of low-frequency ($\sim\mu$eV) penetration depth collected in Fig.~\ref{expfit}.

Finally, a $T^3$ thermal depletion of the thermally averaged superfluid density then straightforwardly follows,
\begin{align}
\label{rho_s_therm}
\frac{\rho_\mathrm{s}}{\rho} = 1 - AT^3,
\end{align}
(with a material dependent coefficient $A$) according to the canonical thermal statistics of $\tilde{N}_0$ in a 3D superfluid with finite stiffness. (c.f. Materials and Methods)
Compared with the $T^4$-depletion of $\rho_\mathrm{s}$ from the linear response~\cite{Ceperley,Landau_vol9},  as expected this $T^3$-depletion of our hydrodynamic $\rho_\mathrm{s}$ is indeed faster (c.f. Discussion) and more appropriate in describing the universal unorthodox thermal fluctuation found in Fig.~\ref{expfit}.
% +-+-+-
%\subsection{Leading order effects of impurities}

Interestingly, upon introducing weak on-site disorder to the bosonic system described above, a perturbative calculation (c.f. SI Appendix, Section 6) shows that the leading low-temperature depletion of the superfluid density increases to $T$-linear. This result is relevant here because such a strong depletion will completely overwhelm the intrinsic $T^3$ behavior at low temperature. Further, many of the modern superconductor families are well-known to exhibit a significant number of impurities, which might explain the observed $T$-linear superfluid density in earlier measurements.

\section*{Discussion}\label{discussion}

\subsection*{Observed \texorpdfstring{$T^3$}{T3}-universality implies bosonic superfluidity}
Remarkably, the above generic $T^3$-depletion of bosonic superfluidity at low-temperature exactly reproduces the unorthodox universal fluctuation identified in Fig.~\ref{expfit} unaccounted for by standard theories of superconductivity.
These modern superconductors therefore must be understood \textit{not} as Cooper-paired Fermi liquids, but rather as charged superfluids of emergent bosons, \textit{regardless} of the microscopic origin of such emergent bosonic carriers.
Furthermore, recall that above the superconducting temperature many of these materials exhibit non-Fermi liquid behavior~\cite{powell2011quantum,borisenko2008pseudogap,chen2017emergence,umemoto2019pseudogap,iwaya2007local,bruin2013similarity,lohneysen1994non,paglione2007incoherent,ironpnictides_1,ironbase_1,fink2015non,dai2015spin,nickelate,kim,valla1999evidence,abdel2006anisotropic,johnston2012evidence,kaminski2005momentum,tao1997observation,renner1998pseudogap,timusk1999pseudogap,lang2002imaging,damascelli2003angle,yang2008emergence,particlehole,pseudogap_t_depen,he2011single,arpes_cuprate,arpes_cuprate_2,he2011single,he2011single}, which is trivially consistent with the fact that a Bose liquid is a non-Fermi liquid~\cite{xander1,xinlei1,Tao1}.
Altogether, our results suggest a new paradigm where the dominant quantum states of these superconducting materials are those of an emergent Bose liquid~\cite{prepairmodelweiku,yildirimWeiku,laurence1,hegg1,xander1,xinlei1} and their low-temperature superconducting phase is just the corresponding superfluid phase.

\subsection*{Prediction: $T^2$-depletion in 2D system}
Furthermore, since superfluidity is known to persist in two dimensions~\cite{BishopReppy2DSFDiscovery}, contrary to the inapplicability of BCS-meanfield treatment below 3D~\cite{Hohenberg-MW}, the above strictly Galilean invariant theory of bosonic superfluidity is naturally capable of explaining pure 2D superconductivity through a finite `true condensate' with stiffness.
Indeed, contrary to the lack of bare condensate in 2D Bose gas~\cite{Pathria}, a finite `true condensate' is expected in a 2D Bose liquid with stiffness.
(In addition, the standard particle-\textit{non}-conserving meanfield argument~\cite{Halperin-HMW} against breaking U(1) symmetry in 2D systems does not directly apply to a particle-conserving Bose liquid.)
Specifically, our strictly Galilean invariant superfluid theory gives a universal $T^2$-depletion of the superfluid density in a pure-2D bosonic superfluid. (c.f. Materials and Methods)
It would be highly valuable to experimentally examine the low-temperature thermal depletion in pure-2D superconducting materials, such as those discovered in the interface~\cite{Bozovic_InterfaceSC}, single-layer FeSe~\cite{Wang2012,Ge2014,Huang2016}, twisted bilayer graphenes~\cite{Cao_TBG_2018,Yankowitz_TBG_2019,Lu_TBG_2019,Saito_TBG_2020,Cao_TBG_strange_2020}, twisted trilayer graphenes~\cite{Chen_ABC1_2019,Chen_ABC2_2019,Cao_TTG_2021,Tsai_TTG_2021}, and few-layer transition metal dichalcogenides~\cite{Wang_2020,Sajadi_2018,Ichinokura_2019,Fu_2017}.

\subsection*{Potential application to \texorpdfstring{$T$}{T}-linear depletion in materials}
Moreover, many unconventional superconductors were found to display $T$-linear depletion of superfluid density, including earlier samples of some materials in Fig.~\ref{expfit} [e.g. in~\cite{HardyetalDirtydWave}].
Interestingly, upon introduction of disorder, the above strictly Galilean invariant superfluid theory switches from $T^3$ to $T$-linear depletion of the low-temperature superfluid density (c.f. SI Appendix, Section 6).
Given that most of the measured samples do contain various amounts of non-negligible dopants, vacancies, dislocations, or impurities, bosonic superfluidity offers an explanation alternative to current proposals~\cite{AGR1991}.
[Of course, such cases should be considered distinct from \textit{another} universality class, whose $T$-linear thermal fluctuation results from gapless amplitude fluctuations in clean systems, for example via nodal Bogoliubov quasi-particle excitation~\cite{nodal_review}.]

\subsection*{Applicability to lattice systems}
Since the above Galilean invariant theory of hydrodynamic superfluidity starts with a low-energy effective description \eqref{bare_H}, its applicability naturally extends to all lattice systems with \textit{emergent} Galilean invariance at long space-time scales.
That is, as long as the lattice effects can be \textit{smoothly} integrated into the renormalization of the low-energy physics belonging to the same fixed point described by \eqref{Hamiltonian} at long spatial scale, the resulting qualitative universal $T^3$ thermal depletion would apply.
On the other hand, it is not expected to apply to systems with strong correlation-induced bifurcation into a different fixed point [having distinct system size-scaling of coefficients in \eqref{Hamiltonian} or with contained field, $\tilde{a}^{\dag}_{\mathbf{k}}$], such as the Mott insulating phase.

\subsection*{Qualitative distinction of hydrodynamic superfluid density from the linear-response one}
%\subsection{Hydrodynamic vs linear-response definitions of superfluid density}
\label{hydrodynamic_vs_LR}
The concept of superfluid density was originally introduced in Landau's two-fluid picture for describing \textit{hydrodynamics} of a superfluid~\cite{landau2013fluid}.
Later, due to easier access in scattering experiments and in quantum theories~\cite{Baym_1968,LeeHuangYang}, superfluid density defined through linear response became widely employed.
However, it is not widely understood that these two definitions account for different components of the fluid and are thus \textit{nonequivalent}, as we will show now.

In hydrodynamics, the explicit inclusion of the environment, say a wall, is \textit{necessary}, to explicitly pin the current of the fluid's normal component through a boundary condition.
Furthermore, each small local component of the fluid is assumed to have \textit{reached} thermal equilibrium with its surroundings.
Therefore, the remaining superfluid component has a stringent condition that it must resist all forms of dissipation throughout the \textit{entire} complex equilibrium process.
Quantum mechanically, this requires a protection against low-velocity excitations in every direction.

On the other hand, linear response of the system is insensitive to the environment.
Correspondingly, the thermal ensemble can be fixed before the weak probing perturbation is applied.
Within such a linear response setup, the superfluid component only needs to retain a curl-less flow \textit{without} having to re-thermalize with the environment.
(Such a \textit{lower} standard is responsible for its resulting artificial finite superfluid density even in a non-superfluid ideal Bose gas.)

Given the drastic distinction between the physical conditions in these two definitions, one would expect their resulting superfluid densities to differ non-trivially.
To see this, it suffices to observe a distinct superfluid density from these two definitions in an arbitrary quantum eigen-state of the Hamiltonian, as a further thermal ensemble average over all eigen-states would naturally retain or even enhance such difference.

\begin{figure}[t!]
\includegraphics[width=8.7cm]{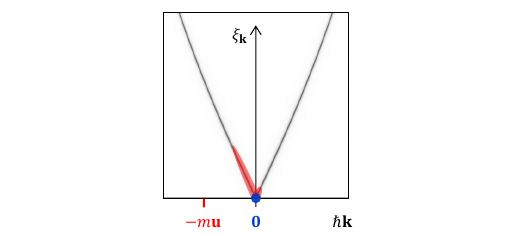}
\centering
\caption{\textbf{Thermal distribution in the condensate frame for the linear-response analysis}
In static linear response analysis, the normal-fluid component of the fluid is assumed to follow a standard thermal distribution in the wall frame of velocity $-\mathbf{u}$ against the condensate (blue dot).
As illustrated by the red shade, the corresponding thermal distribution in the condensate frame thus takes the boosted form~\cite{Landau_vol9}, $n_\mathrm{B}(\xi_\mathbf{k}+\mathbf{u}\cdot\hbar\mathbf{k})$, of the Bose-Einstein distribution function.
An eigen-state of the system with momentum occupation of eigen-particles resembling this distribution is therefore a good representative quantum pure state that gives distinct hydrodynamic and linear-response superfluid density.
}
\label{fig_inconsistency}
\end{figure}

Consider for example a representative eigen-state of $H_{\text{diag}}$ shown in Fig.~\ref{fig_inconsistency} in the condensate frame.
Notice that some eigen-particles are with finite momentum and thus outside the true condensate, as a result of coupling, $\mathbf{u}\cdot\mathbf{P}\equiv\mathbf{u}\cdot\sum_\mathrm{k}\hbar\mathbf{k} a^{\dagger}_{\mathbf{k}} a_{\mathbf{k}}$, to a weak uniform external velocity field $\mathbf{u}$.
Under the linear response (LR) definition, this state gives a normal fluid density,
\begin{align}
\label{rho_n_LR}
\rho^\mathrm{(LR)}_\mathrm{n}=\lim_{\mathbf{u}\to\mathbf{0}}\frac{1}{\mathcal{V}}\frac{\mathbf{u}}{u^2}\cdot\mathbf{P},
\end{align}
from which the superfluid density, $\rho^\mathrm{(LR)}_\mathrm{s}\equiv \rho -\rho^\mathrm{(LR)}_\mathrm{n}$, can be inferred.
Importantly, as a standard property of linear response~\cite{Kubo_1957}, such static response can be directly computed via the intrinsic retarded momentum-momentum correlation function~\cite{Baym_1968,Ceperley} in the \textit{absence} of the $\mathbf{u}$ field~\cite{Baym_1968,Ceperley}.

On the other hand, in the hydrodynamic description of the two-fluid model, the superfluid density corresponds to the mass density of the entropy-free, boundary-insensitive persisting flow described by a curl-less velocity field.
Obviously, for the representative eigen-state shown in Fig.~\ref{fig_inconsistency}, only eigen-particles in the \textit{single} momentum $\hbar\mathbf{k}=\mathbf{0}$ (the true condensate) are protected from the boundary induced thermalization process and contribute to such an entropy-free component.
Given that at long wavelength eigen-particles carry the inertial mass of bare particles (c.f. Materials and Methods), the hydrodynamic definition of superfluid density is thus, $m\tilde{N}_\mathrm{0}/\mathcal{V}$, as already established above in \eqref{superfluid_density}.

Notice that the hydrodynamic normal-fluid density, \eqref{normal_fluid_density},
\begin{align}
\label{rho_n_HD}
\rho_\mathrm{n}=m\frac{\tilde{N}_\mathrm{n}}{\mathcal{V}}=\frac{1}{\mathcal{V}}\frac{\mathbf{v}_\mathrm{n}}{v^2_\mathrm{n}}\cdot\mathbf{P},
\end{align}
corresponds to the average flow velocity of these uncondensed eigen-particles, $\mathbf{v}_\mathrm{n}\equiv\sum_\mathrm{k\neq 0}\frac{\hbar\mathbf{k}}{m} a^{\dagger}_{\mathbf{k}} a_{\mathbf{k}}/\tilde{N}_\mathrm{n}$, \textit{not} that of the wall, $-\mathbf{u}$, in great contrast to \eqref{rho_n_LR}.
Therefore, the hydrodynamic superfluid density is, \textit{by definition}, different from the linear response one.
Furthermore, as can be visualized from Fig.~\ref{fig_inconsistency}, it is straightforward to verify that in the low-temperature limit, generally $v \ll u$ and thus $\rho_\mathrm{n} > \rho^\mathrm{(LR)}_\mathrm{n}$, as expected from the more stringent conditions the hydrodynamic superfluid has to satisfy.
Correspondingly, at low temperature, $\rho_\mathrm{s} < \rho^\mathrm{(LR)}_\mathrm{s}$, consistent with the faster $T^3$-depletion of the former than the $T^4$ of the latter.

Given the general distinction of these two inconsistent definitions of superfluid density, one must exercise caution in comparing results from different types of experiments or theories.
Particularly, results reported in Fig.~\ref{expfit} are from penetration depth of extremely long time scale (\textit{low}-frequency), well within the realm of hydrodynamic superfluidity.
One thus expects a $T^3$-depletion as found in the above Galilean invariant theory of hydrodynamic superfluidity, instead of a $T^4$-depletion~\cite{HuangYang,LeeHuangYang} from the linear response of the system.

\subsection*{Key improvements over previous theories}
Our theory owes its ability in obtaining the previously inaccessible hydrodynamic superfluid density to three significant improvements over previous theories.
First, our derivation strictly enforces particle conservation.
While the previous Bogoliubov meanfield treatment might be reasonable to realize the phase transition, the lack of associated continuity equation surely renders transport properties highly uncontrolled.
Second, our microscopic supercurrent is properly defined directly through a true condensate of long-lived eigen-particles.
In contrast, the standard definition is indirectly associated with a bare condensate $N_0(t)$ that fluctuates in time and does not account for the full superflow~\cite{HuangYang,LeeHuangYang}.
Third and most importantly, the essential inertial mass flow velocity $\mathbf{v}_\mathrm{I}\equiv\frac{\hbar\mathbf{k}}{m}$, \textit{absent} in previous treatments~\cite{Landau_vol9,HuangYang,LeeHuangYang}, is explicitly included in our Galilean invariant current fluctuation $\tilde{\mathbf{J}}_\mathbf{q}$.
As illustrated in Materials and Methods, the dominant vibration velocity $v_\mathrm{v}\hat{\mathbf{q}}$ in previous studies describes \textit{only} the propagation of the vibration unrelated to the inertial mass flow.
Indeed, careful hydrodynamic analysis~\cite{prix} has shown that sound-based evaluation of normal fluid density~\cite{Landau_vol9,HuangYang,LeeHuangYang} only accounts for the ``entropy flow'', not the essential ``mass flow''.

\subsection*{Second sound}
The vibrational modes' insensitivity to the inertial mass flow has another serious implication on the existing experimental observation of the thermal depletion of superfluid density.
In great contrast to the measurement of bare condensate density~\cite{Harling_1971} and excitation spectrum~\cite{Yarnell_1959} in liquid helium, to the best of our knowledge, nearly all observations of the superfluid density below 1K are obtained via the second sound (temperature wave) measurement~\cite{second_sound_fit, second_sound_exp1}.
However, since the propagation of the entropy-associated low-energy elementary excitations is completely overwhelmed by the vibration velocity $\mathbf{v}_\mathrm{v}$ rather than the inertial mass velocity $\mathbf{v}_\mathrm{I}$ [c.f. Fig.~\ref{fig_supercurrent}(a)], it is unreasonable to expect properties of second sound to provide a sensitive measure of the inertial mass-carrying superfluid density.

Indeed, it was shown long ago~\cite{Ward1951,Ward1952,phonon_transport} that at low temperature the propagation of the vibrational modes alone accurately describes the second sound, in the \textit{absence} of the superfluidity.
Even in solids that are not superfluid, second sound with similar temperature dependence to liquid helium has been experimentally observed~\cite{secondsound1,secondsound2,secondsound3}.
In fact, given that $\mathbf{v}_\mathrm{v}$ is temperature-insensitive and the low-temperature elementary excitations are dominated by it, our more complete description would fall back to the standard theory of second sound and reproduce such experiments.
Therefore, it would be highly valuable if other techniques sensitive to the real mass flow~\cite{wire,stackofdisks} can be applied to low enough temperature to directly examine the thermal depletion of the superfluid density.
On the other hand, if indeed most modern superconductors in Fig.~\ref{expfit} are simply charged superfluids of emergent bosons, the sensitive response to electromagnetic fields of these charged bosons would offer an unsurpassed advantage over charge-neutral atoms in studying low-temperature superfluid properties.

\section*{Conclusion}\label{conclusion}

In conclusion, we identify from published experimental data a \textit{universal} $T^3$ thermal depletion of the low-temperature 3D superfluid density in nearly all modern superconductor families.
Contrary to standard lore, this unorthodox behavior persists \textit{independently} of their (fully gapped or nodal) superconducting gap structure.
Absent any known generic theory to account for this, we develop a strictly Galilean invariant microscopic quantum theory of superfluidity based on long-lived eigen-particles.
With strict particle conservation and inclusion of necessary inertial mass velocity in the theory, the superfluid density is then proven rigorously equivalent to the `true condensation' density for any quantum state of interacting bosonic systems, thus naturally producing the identified universal $T^3$ thermal fluctuation.
Altogether, the revealed fluctuation inevitably leaves charged superfluid of emergent bosons as the only remaining theory capable of describing the quantum nature of the unconventional superconductivity in these materials.

\matmethods{
\label{materials-and-methods}
\subsection*{Fitting procedure for experimental data collapse}
\label{fitpro}

\begin{figure}[t!]
        \centering
    \includegraphics[width=8.7cm]{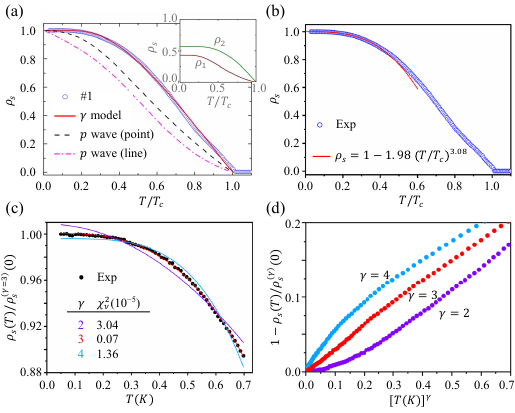}
	\caption{\textbf{Demonstration of the fitting procedure} applied to (a) the superfluid density $\rho_\mathrm{s}$ vs  reduced temperature $T/T_c$ in a data set~\cite{7LaNiGa2} for $\rm{LaNiGa_2}$. (b) Extracted data (blue) from (a) with best-fit power-law line (red) fit to data below a threshold where the fit converges. (c) A simple comparison between least squares fits to $\rho_\mathrm{s}(T)=\rho_\mathrm{s}^{(\gamma)}(0)(1-AT^\gamma)$ with fixed integer exponents $\gamma=2$ (purple), 3 (red), and 4 (blue) of the data set in (b) with standard reduced chi-square deviation $\chi_\nu^2$ shown in the table. Notice the very poor fit with $\gamma=2$ and $\gamma=4$ with opposite data clustering above and below the fit line, indicating that the best fit lies between them. (d) Plot of $1-\rho_\mathrm{s}(T)/\rho^{(\gamma)}_\mathrm{s}(0)$ vs $T^\gamma$ clearly illustrating that only $\gamma=3$ fits well (i.e. converges to a straight line) at low temperature.}
	\label{example}
\end{figure}

\begin{figure*}[t!]
        \centering
    \includegraphics[width=17.8cm]{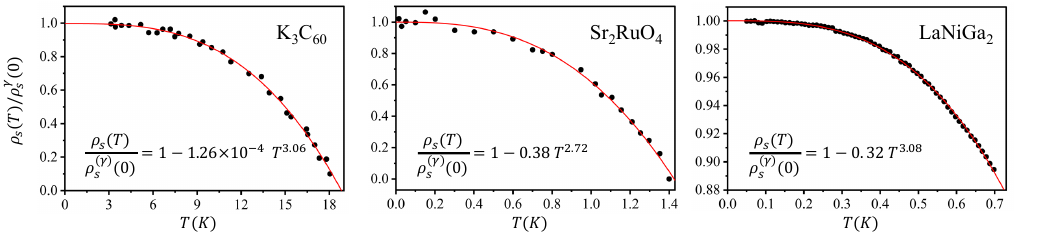}
	\caption{\textbf{Examples of $\rho_\mathrm{s}^{(\gamma)}(0)(1-AT^\gamma)$ fits with non-integer} $\gamma$ \textbf{treated as a free fitting parameter} to the experimental data~\cite{7LaNiGa2,12K3C60,19Sr2RuO4} of three materials, all giving an exponent $\gamma$ closest to integer 3.}
	\label{Bestfit}
\end{figure*}

In this section we describe the procedure used to generate the data collapse plots in Fig.~\ref{expfit}. For clarity, we illustrate the procedure for $\rm{LaNiGa_2}$ as an example. A figure taken from Ref.~\cite{7LaNiGa2} is displayed in Fig.~\ref{example}(a). Using the Origin Digitizer, we extract the superfluid density data displayed in Fig.~\ref{example}(b) and fit the extracted data to a single power-law curve of the form 
\begin{equation}
\label{TheFit}
\rho_\mathrm{s}(T)=\rho_\mathrm{s}^{(\gamma)}(0)(1-AT^\gamma).
\end{equation}
Here, $\rho_\mathrm{s}^{(\gamma)}(0)$, $\gamma$, and $A$ are fitting parameters, with $\rho_\mathrm{s}^{(\gamma)}(0)$ denoting the zero-temperature value extrapolated from the corresponding power-law fit.

Since we are interested in the low-temperature behavior of the superfluid density in these materials, we discard data near the transition temperature systematically until we find that the fitting procedure appears to converge. The result of such a converged fitting is displayed as a red dotted line in Fig.~\ref{example}(b). In fact, all of the materials presented here exhibit a resulting power-law fitting with $\gamma\sim3$, and to illustrate this fact we present three such fittings side-by-side in Fig.~\ref{Bestfit}.

We are interested in the leading order fluctuations at the lowest temperatures and far from any phase transition. A cursory attempt to fit to a (gapped) exponential indicates it is an extremely poor fit, so we search for integer power-law monomial behavior. To that end, we simply display each data set along with best fit lines for fixed integer exponents ($\gamma=2,3,4$ corresponding to purple, red, and blue dotted lines respectively) in Fig.~\ref{example}(c), where it is clear that $\gamma=2$ and $\gamma=4$ each provide very poor fits that significantly underestimate and overestimate $\gamma$ respectively. To see the convergence to $\gamma=3$ more clearly in this example, we plot $1-\rho_\mathrm{s}(T)/\rho_\mathrm{s}^{(\gamma)}(0)$ vs $T^\gamma$ in Fig.~\ref{example}(d) to show that the data converge to a straight line (red points) for $\gamma=3$ but not for $\gamma=2$ (purple) or $\gamma=4$ (blue). In Fig.~\ref{fitset} we display the integer exponent fitting, analogous to Fig.~\ref{example}(c), for each data set presented in the main text.

In order to display the results in a single data-collapse figure, especially since the onset temperatures of this effect ($\sim 0.14$K$-$$18.05$K) span more than two orders of magnitude across all of these materials, we further rescale the temperature to absorb the coefficient
\begin{align}
\label{rescaling}
\tilde{T} &= A^{-\frac{1}{3}}.
\end{align}

Finally, in the inset of Fig.~\ref{expfit}, we illustrate the convergence of all data sets to $\gamma=3$ via a plot analogous to the example in Fig.~\ref{example}(d). For clarity, we introduce a data-set dependent `zoom' factor $Z$, identified for each material in Table~\ref{zoom}, such that the deviation from $\gamma=3$ occurs at around the same point in the inset.

\begin{table}[t!]
\centering
\caption{\textbf{The zoom factor} $Z$ used to scale the inset of Fig.~\ref{expfit} for easier visualization.}
\label{zoom}
\resizebox{0.85\columnwidth}{!}{
\begin{tabular}[t]{lcc}
\hline
$\mathrm{Material}$&$1/Z$\\
\hline
$\mathrm{MgB_2}$~\cite{2MgB2_2}  & 0.07\\
$\mathrm{CsV_3Sb_5}$~\cite{3CsV3Sb5} & 0.15\\
$\mathrm{LiFeAs}$~\cite{4LiFeAs} & 0.13 \\
$\mathrm{NbSe_{2}}$~\cite{5NbSe2}  &0.1 \\
$\mathrm{LaNiGa_2}$~\cite{7LaNiGa2} & 0.25\\
$\mathrm{Mo_5PB_2}$~\cite{18Mo5PB2}& 0.75\\
$\mathrm{PrPt_4Ge_{12}}$~\cite{8PrPt4Ce14} &0.18\\
$\mathrm{PdTe_2}$~\cite{11PbTe2}  & 0.6\\
$\mathrm{K_3C_{60}}$~\cite{12K3C60} & 1\\
$\mathrm{SrPd_2Ge_2}$~\cite{14SrPd2Ge2} & 0.5 \\
$\mathrm{Pr_{2-x}Ce_xCuO_{4-\delta}}$(x = 0.131)~\cite{9PCCO} & 0.08 \\
$\mathrm{Ba_{0.6}K_{0.4}BiO_3}$~\cite{17BKBO}  &0.45\\
$\mathrm{LaMo_6Se_8}$~\cite{17BKBO} & 0.45\\
$\mathrm{Sr_2RuO_4}$~\cite{19Sr2RuO4}   & 0.7 \\
$\mathrm{UTe_2}$~\cite{20UTe2} & 0.015 \\
$\mathrm{YBa_2Cu_3O_{6+y}}$~\cite{hardyUDCuprate} & 0.1 \\
$\mathrm{Nb:SrTiO_3}$~\cite{21STO} & 1 \\
$\kappa$-$\mathrm{(BEDT-TTF)_2Cu[N(CN)_2]Br}$~\cite{15Organic}& 0.2 \\
\end{tabular}
}
\end{table}

\begin{SCfigure*}[\sidecaptionrelwidth][t!]
        \centering
\includegraphics[width=11.4cm]{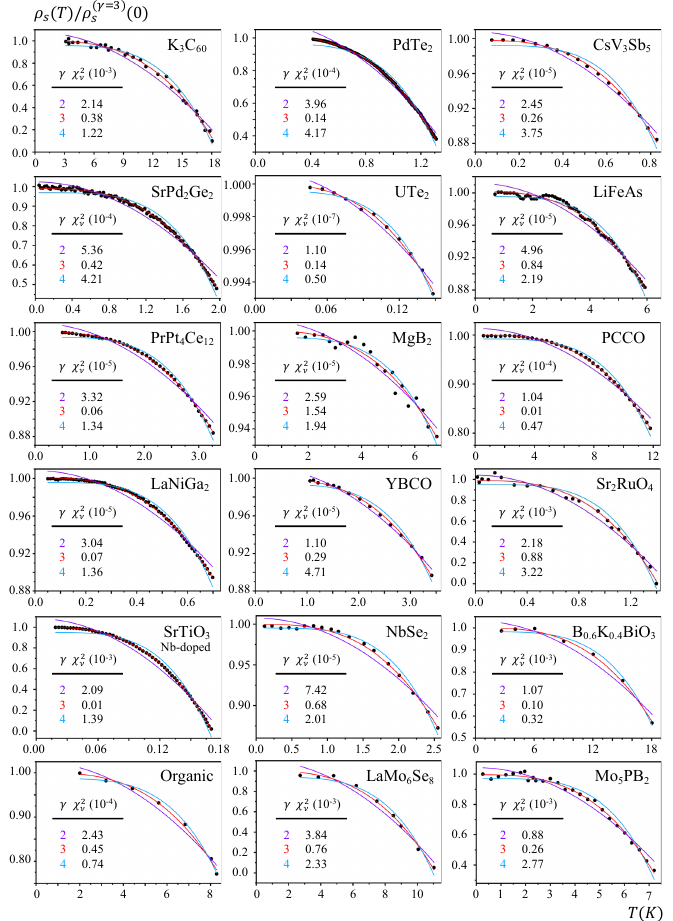}
	\caption{\textbf{Least squares fits to} $\rho_\mathrm{s}^{(\gamma)}(0)(1-AT^\gamma)$ \textbf{with fixed integer} $\gamma = 2$ (purple), $3$ (red), and $4$ (blue) for several example data sets at low temperature, with standard reduced chi-square deviation $\chi_\nu^2$ given in the table. In all cases, $\gamma=3$ is superior to $2$ and $4$.}
	\label{fitset}
\end{SCfigure*}

\subsection*{Eigen-particles' preservation of conserved quantum number}
\label{quantum_number_preservation}

For long-wavelength properties the eigen-particles carry the same conserved quantum numbers as the bare particles, such as inertial mass $m$, charge, and momentum $\hbar\textbf{k}$ of the bare particles.
This is evident since the unitary transformation $U$ must respect all constants of motion, such as total particle number $N$, inertial mass $M$, and kinematic momentum $\mathbf{P}$, $[U,N]=[U,M]=[U,\mathbf{P}]=0$.
One thus finds,
\begin{equation}
    N\equiv\sum_\mathbf{k}a^\dagger_\mathbf{k}a_\mathbf{k}=U^\dag NU=\sum_\mathbf{k}\tilde{a}^\dagger_\mathbf{k}\tilde{a}_\mathbf{k},
\end{equation}
\begin{equation}
    M\equiv\sum_\mathbf{k}ma^\dagger_\mathbf{k}a_\mathbf{k}=U^\dag MU=\sum_\mathbf{k}m\tilde{a}^\dagger_\mathbf{k}\tilde{a}_\mathbf{k}\equiv \tilde{M},
\end{equation}
\begin{equation}
    \mathbf{P}\equiv\sum_\mathbf{k}\hbar\mathbf{k}a^\dagger_\mathbf{k}a_\mathbf{k}=U^\dag \mathbf{P}U=\sum_\mathbf{k}\hbar\mathbf{k}\tilde{a}^\dagger_\mathbf{k}\tilde{a}_\mathbf{k}\equiv\tilde{\mathbf{P}}.
\end{equation}
Superior to the generally fluctuating bare particles, the eigen-particles are therefore ideal when number-counting for properties of a long space-time scale.

\subsection*{Galilean symmetry of $\epsilon_{\mathbf{k}_1,\mathbf{k}_2,\cdots}$}
    For a Galilean invariant system moving with velocity $\mathbf{u}$ relative to the lab frame, the Hamiltonian observed in this frame is given by
    \begin{align}
        H' = H + \mathbf{u}\cdot \mathbf{P} + \frac{1}{2} Mu^2, \label{GalileanBoost}
    \end{align}
    where $H$ and $\mathbf{P}$ are quantities defined in the system's rest frame.
    In the diagonal-representation, the Galilean boost of \eqref{Hamiltonian} can be expressed as 
    \begin{align}
        &H'_\mathrm{diag} = \sum_\mathbf{k} \epsilon_{\mathbf{k}+\mathbf{k}_0} \tilde{a}_\mathbf{k}^\dagger \tilde{a}_\mathbf{k}+\frac{1}{2!}\sum_{\mathbf{k}_1,\mathbf{k}_2} \epsilon_{\mathbf{k}_1+\mathbf{k}_0,\mathbf{k}_2+\mathbf{k}_0}\tilde{a}^\dagger_{\mathbf{k}_1}\tilde{a}^\dagger_{\mathbf{k}_2}\tilde{a}_{\mathbf{k}_2}\tilde{a}_{\mathbf{k}_1}\nonumber\\
        &+\cdots + \frac{1}{N!}\sum_{\mathbf{k}_1\cdots\mathbf{k}_N}\epsilon_{\mathbf{k}_1+\mathbf{k}_0,\cdots,\mathbf{k}_N+\mathbf{k}_0} \tilde{a}^{\dagger}_{\mathbf{k}_1} \cdots \tilde{a}^{\dagger}_{\mathbf{k}_N} \tilde{a}_{\mathbf{k}_N} \cdots \tilde{a}_{\mathbf{k}_1},
    \end{align}
    where $m$ is the $\mathbf{k}$-independent single-particle mass and $\mathbf{k}_0 \equiv m\mathbf{u}/\hbar$. Note that for an arbitrary $\mathbf{u}$, the one-body term 
    \begin{align}
        \sum_\mathbf{k} \epsilon_{\mathbf{k}+\mathbf{k}_0} \tilde{a}_\mathbf{k}^\dagger \tilde{a}_\mathbf{k} =& \sum_\mathbf{k} \frac{\hbar^2(\mathbf{k}+m\mathbf{u}/\hbar)^2}{2m} \tilde{a}_\mathbf{k}^\dagger \tilde{a}_\mathbf{k} \nonumber\\=& \sum_\mathbf{k} \epsilon_\mathbf{k} \tilde{a}_\mathbf{k}^\dagger \tilde{a}_\mathbf{k}+\mathbf{u}\cdot \tilde{\mathbf{P}}+\frac{1}{2}\tilde{M}u^2
    \end{align}
    already fully incorporates the boosted terms on the right-hand side of \eqref{GalileanBoost}, since $\tilde{M} = M$ and $\tilde{\mathbf{P}}=\mathbf{P}$. Thus, the multi-body terms must remain invariant under the Galilean boost, which implies
    \begin{equation}
    \epsilon_{\mathbf{k}_1+\mathbf{k}_0,\mathbf{k}_2+\mathbf{k}_0,\cdots} = \epsilon_{\mathbf{k}_1,\mathbf{k}_2,\cdots}, \quad \forall~ \mathbf{k}_0.
    \label{Galilean_symmetry}
    \end{equation}

\subsection*{Eigen-particle removal energy and one-body propagator}
\label{phase_stiffness}

The one-body propagator of the eigen-particles is defined in the standard form of a time-ordered Green's function:
\begin{align}
\label{To-Gf1}
\tilde{G}(\mathbf{k},t,t')
&\equiv
-iT_t
\tilde{a}_{\mathbf{k}}(t)
\tilde{a}_{\mathbf{k}}^\dagger(t')\nonumber
\\
&=
-i\Big[
\theta(t'-t)
\tilde{a}_{\mathbf{k}}^\dagger(t')
\tilde{a}_{\mathbf{k}}(t)
+
\theta(t-t')
\tilde{a}_{\mathbf{k}}(t)
\tilde{a}_{\mathbf{k}}^\dagger(t')
\Big].
\end{align}

The commutator
$
[H,\tilde{a}_{\mathbf{k}}^\dagger]
\equiv
\tilde{a}_{\mathbf{k}}^\dagger
\tilde{\epsilon}_{\mathbf{k}}
$
defines the diagonal operator $\tilde{\epsilon}_{\mathbf{k}}$, which can be expanded as
\begin{align}
\tilde{\epsilon}_{\mathbf{k}}
&=
\epsilon_{\mathbf{k}}
+
\sum_{\mathbf{k}_1}
\epsilon_{\mathbf{k}\mathbf{k}_1}
\tilde{a}_{\mathbf{k}_1}^\dagger
\tilde{a}_{\mathbf{k}_1}
+
\frac{1}{2!}
\sum_{\mathbf{k}_1\mathbf{k}_2}
\epsilon_{\mathbf{k}\mathbf{k}_1\mathbf{k}_2}
\tilde{a}_{\mathbf{k}_1}^\dagger
\tilde{a}_{\mathbf{k}_2}^\dagger
\tilde{a}_{\mathbf{k}_2}
\tilde{a}_{\mathbf{k}_1}
+\cdots
\nonumber\\
&\equiv
\epsilon_{\mathbf{k}}+\tilde{\Sigma}_{\mathbf{k}}
=
\frac{\hbar^2\mathbf{k}^2}{2m}
+\tilde{\Sigma}_{\mathbf{k}}.
\end{align}
It follows that
$
\tilde{a}_{\mathbf{k}}^\dagger(t)
=
\tilde{a}_{\mathbf{k}}^\dagger
e^{it\tilde{\epsilon}_{\mathbf{k}}}.
$
The Fourier transform of the Green's function is then
\begin{align}
\label{TO-Gf3}
\tilde{G}(\mathbf{k},\omega)
&\equiv
\int d(t-t')\,
\tilde{G}(\mathbf{k},t,t')
e^{i\omega(t-t')}
\nonumber\\
&=
-\tilde{a}_{\mathbf{k}}^\dagger
\frac{1}
{\omega-\tilde{\epsilon}_{\mathbf{k}}-i\delta}
\tilde{a}_{\mathbf{k}}
+
\frac{1}
{\omega-\tilde{\epsilon}_{\mathbf{k}}+i\delta}
\tilde{a}_{\mathbf{k}}
\tilde{a}_{\mathbf{k}}^\dagger.
\end{align}
The first term describes the removal of an eigen-particle from the system, whereas the second describes the addition of an eigen-particle.
It is therefore useful to define the eigen-particle removal energy as
\begin{equation}
\xi_{\mathbf{k}}
\equiv
\left\langle
\frac{1}{\tilde{N}_{\mathbf{k}}}
\tilde{a}_{\mathbf{k}}^\dagger
\tilde{\epsilon}_{\mathbf{k}}
\tilde{a}_{\mathbf{k}}
\right\rangle,
\end{equation}
which characterizes the removal pole of
$\tilde{G}(\mathbf{k},\omega)$.

\subsection*{Dressed current}
\label{dressed_current_section}

Because the eigen-particles carry the same inertial mass $m$ and momentum $\hbar\mathbf{k}$ as the bare particles, these long-lived particles provide a convenient framework for tracking the mass flow of the system over long length and time scales. Specifically, the eigen-particle current in the Heisenberg picture is defined as the time derivative of the dressed polarization operator,
\begin{equation}
\label{AdressedJ}
\tilde{\mathbf{J}}_{\mathbf{q}}
\equiv
\frac{\mathrm{d}\tilde{\mathbf{\Pi}}_{\mathbf{q}}}{\mathrm{d}t}
=
\frac{1}{i\hbar}
\left[
\tilde{\mathbf{\Pi}}_{\mathbf{q}},H
\right].
\end{equation}
Here, the dressed polarization operator at wave vector $\mathbf{q}$ is
\begin{equation}
\tilde{\mathbf{\Pi}}_{\mathbf{q}}
\equiv
\frac{1}{\mathcal{V}}
\sum_{\mathbf{R}}
e^{i\mathbf{q}\cdot\mathbf{R}}
\mathbf{R}\,
\tilde{a}_{\mathbf{R}}^\dagger
\tilde{a}_{\mathbf{R}},
\end{equation}
where $\tilde{a}_{\mathbf{R}}^\dagger$, obtained by Fourier transforming
$\tilde{a}_{\mathbf{k}}^\dagger$, creates an eigen-particle centered at
$\mathbf{R}$.
This definition ensures that the continuity equation
\begin{equation}
\frac{\mathrm{d}\tilde{n}_{\mathbf{q}}}{\mathrm{d}t}
-
i\mathbf{q}\cdot\tilde{\mathbf{J}}_{\mathbf{q}}
=
0
\end{equation}
is satisfied on length scales $2\pi/|\mathbf{q}|$ larger than the characteristic size of the eigen-particles, where
\begin{equation}
\tilde{n}_{\mathbf{q}}=
\frac{1}{\mathcal{V}}
\sum_{\mathbf{k}}
\tilde{a}_{\mathbf{k}+\mathbf{q}}^\dagger
\tilde{a}_{\mathbf{k}}
\end{equation}
is the eigen-particle density operator.

The diagonal form of \eqref{Hamiltonian} yields a particularly simple expression for the dressed current,
\begin{align}
\tilde{\mathbf{J}}_{\mathbf{q}}
&=
\frac{1}{\hbar\mathcal{V}}
\sum_{\mathbf{k}}
\tilde{a}_{\mathbf{k}+\mathbf{q}}^\dagger
\mathbf{\nabla}_{\mathbf{q}}
\left(
\tilde{\epsilon}_{\mathbf{k}+\mathbf{q}}
-
\tilde{\epsilon}_{\mathbf{k}}
\right)
\tilde{a}_{\mathbf{k}}
\nonumber\\
&\quad+
\frac{1}{\hbar\mathcal{V}}
\sum_{\mathbf{k}}
\left(
\mathbf{\nabla}_{\mathbf{q}}
\tilde{a}_{\mathbf{k}+\mathbf{q}}^\dagger
\right)
\left(
\tilde{\epsilon}_{\mathbf{k}+\mathbf{q}}
-
\tilde{\epsilon}_{\mathbf{k}}
\right)
\tilde{a}_{\mathbf{k}}
\nonumber\\
&\xrightarrow{\mathbf{q}\to\mathbf{0}}
\frac{1}{\mathcal{V}}
\sum_{\mathbf{k}}
\tilde{a}_{\mathbf{k}+\mathbf{q}}^\dagger
\tilde{\mathbf{v}}_{\mathbf{k}+\mathbf{q}/2}
\tilde{a}_{\mathbf{k}},
\label{Adressed_current}
\end{align}
where the diagonal eigen-particle velocity operator is defined by
\begin{equation}
\tilde{\mathbf{v}}_{\mathbf{k}}
\equiv
\frac{1}{\hbar}
\mathbf{\nabla}_{\mathbf{k}}
\tilde{\epsilon}_{\mathbf{k}}
\nonumber
=
\frac{1}{\hbar}
\mathbf{\nabla}_{\mathbf{k}}
\left(
\epsilon_{\mathbf{k}}
+
\tilde{\Sigma}_{\mathbf{k}}
\right)
=
\frac{\hbar\mathbf{k}}{m}
+
\frac{1}{\hbar}
\mathbf{\nabla}_{\mathbf{k}}
\tilde{\Sigma}_{\mathbf{k}}.
\end{equation}
Its expectation value for an occupied eigen-particle state defines the eigen-particle velocity,
\begin{equation}
\boldsymbol{\nu}_{\mathbf{k}}
\equiv
\left\langle
\frac{1}{\tilde{N}_{\mathbf{k}}}
\tilde{a}_{\mathbf{k}}^\dagger
\tilde{\mathbf{v}}_{\mathbf{k}}
\tilde{a}_{\mathbf{k}}
\right\rangle
=
\frac{1}{\hbar}
\mathbf{\nabla}_{\mathbf{k}}
\xi_{\mathbf{k}},
\end{equation}
where $\xi_{\mathbf{k}}$ is the energy required to remove an eigen-particle of momentum $\mathbf{k}$ from the $N$-particle system.

In addition, \eqref{Galilean_symmetry}, which follows from Galilean invariance, further simplifies the uniform current,
\begin{equation}
\tilde{\mathbf{J}}
\equiv
\tilde{\mathbf{J}}_{\mathbf{q}=\mathbf{0}}
=
\frac{1}{\mathcal{V}}
\sum_{\mathbf{k}}
\tilde{a}_{\mathbf{k}}^\dagger
\tilde{\mathbf{v}}_{\mathbf{k}}
\tilde{a}_{\mathbf{k}}
=
\frac{1}{\mathcal{V}}
\sum_{\mathbf{k}}
\frac{\hbar\mathbf{k}}{m}
\tilde{a}_{\mathbf{k}}^\dagger
\tilde{a}_{\mathbf{k}}.
\end{equation}
This result is consistent with the fact that, in a Galilean-invariant system, many-body interactions preserve the total kinematic momentum
$\tilde{\mathbf{P}}$. Consequently,
\begin{equation}
\tilde{\mathbf{J}}
=
\frac{\tilde{\mathbf{P}}}{m\mathcal{V}}
=
\frac{1}{\mathcal{V}}
\sum_{\mathbf{k}}
\frac{\hbar\mathbf{k}}{m}
\tilde{a}_{\mathbf{k}}^\dagger
\tilde{a}_{\mathbf{k}},
\end{equation}
so that the uniform mass current is determined entirely by the total kinematic momentum carried by the inertial mass.

\subsection*{Two physical velocities of current fluctuation in a superfluid}
\label{dressedJ}

\begin{figure}[t!]
        \centering
     \includegraphics[width=8.7cm]{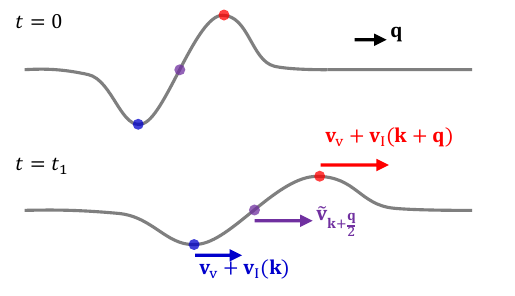}
	\caption{
 \textbf{Illustration of two propagation velocities of current fluctuation in a superfluid}.
 The \textit{longitudinal} propagation of current fluctuation in a superfluid contains two distinct components of velocity: 1) the inertial mass velocity $\mathbf{v}_\mathrm{I}\equiv\frac{\hbar\mathbf{k}}{m}$, and the vibration velocity $\mathbf{v}_\mathrm{v}=v_\mathrm{v}\hat{\mathbf{k}}$.
 For example, $\frac{1}{\mathcal{V}}\tilde{a}^{\dag}_{\mathbf{k}+\mathbf{q}} \tilde{\mathbf{v}}_{\mathbf{k}+\frac{\mathbf{q}}{2}} \tilde{a}_{\mathbf{k}}$ in the current fluctuation $\tilde{\mathbf{J}}_\mathbf{q}$ contributes to a propagating wave packet containing a valley (hole, in blue) with velocity $\mathbf{v}_\mathrm{v}+\frac{\hbar\mathbf{k}}{m}$ and a peak (particle, in red) with a faster velocity $\mathbf{v}_\mathrm{v}+\frac{\hbar}{m}(\mathbf{k}+\mathbf{q})$.
 Over time, the inertial mass associated with the peak would move away from the valley with a relative velocity $\mathbf{v}_\mathrm{I}(\mathbf{q})=\frac{\hbar\mathbf{q}}{m}$.
 The eigen-particle velocity $\tilde{\mathbf{v}}_{\mathbf{k}+\frac{\mathbf{q}}{2}}$ encodes their average velocity $\mathbf{v}_\mathrm{v}+\frac{\hbar}{m}(\mathbf{k}+\frac{\mathbf{q}}{2})$ as the velocity of the node (in purple).}
\label{mass_velocity}
\end{figure}

It is well-established that a superfluid develops a stiffness that gives rise to well-defined vibrations with energy $v_\mathrm{v}\hbar|\mathbf{k}|$ and isotropic propagating velocity $\mathbf{v}_\mathrm{v}\equiv v_\mathrm{v}\hat{\mathbf{k}}$ in its rest frame.
However, this velocity, while dominant near the condensate, does not carry the inertial mass.
This is directly analogous to the lattice of atoms in crystals whose momentum-dependent acoustic vibration (the sound) takes the same form $v_\mathrm{v}\hbar|\mathbf{k}|$ exactly (without higher order correction).
This ensures that the shape of the vibration remains unchanged during propagation, such that only the vibration can propagate but not the atoms themselves.

The superflow, however, refers to the actual flow of inertial mass, as in regular fluid.
Therefore, it is necessary to further include the Galilean invariant kinetic energy $\epsilon_{\mathbf{k}}=\frac{\hbar^2}{2m}\mathbf{k}\cdot\mathbf{k}$ and the corresponding inertial mass velocity $\mathbf{v}_\mathrm{I}\equiv\frac{\hbar\mathbf{k}}{m}$ into consideration.
Even though near the condensate momentum, the inertial mass velocity is sub-leading to the vibration velocity, it is the one that carries the essential information of inertial mass velocity.
Figure~\ref{mass_velocity} gives a simple illustration of how the medium in a superfluid evolves over time that reflects these two components of velocity in the current fluctuation (as a particle-hole excitation in second quantized picture).

It is straightforward to show that only with inclusion of the kinetic contribution, can such a two-component velocity retain the equivalence between the canonical momentum of the fluctuation and the kinematic momentum of the medium.
Furthermore, it is important to note that this form of velocity, $\frac{\hbar\mathbf{k}}{m}+v_\mathrm{v}\hat{\mathbf{k}}$, is already the most general form of velocity in Galilean invariant systems, as any additional contribution would necessarily violate the standard Galilean boost of velocity.

On that note, since the standard Bogoliubov treatment~\cite{Bogoliubov,S-B-PSFText} does not retain this necessary kinematic velocity, it is not too surprising that such a treatment does not uphold the essential $f$-sum rule in its corresponding current response~\cite{HohenbergMartin}.

\subsection*{Thermal depletion}
\label{depletion}

Having established the equivalence between the hydrodynamic superfluid density and the true condensate density, $\rho_\mathrm{s}/\rho=\tilde{a}^{\dag}_{\mathbf{k}_0} \tilde{a}_{\mathbf{k}_0}/N$, it is straightforward to evaluate the temperature dependence of the former from that of the latter.
Specifically, under an extremely weak experimental perturbation, the system is basically in standard thermal equilibrium.

Therefore, given the well-established linear dispersion in the elementary excitations of a $d$-dimensional ($d\ge2$) superfluid~\cite{LandauSF2Long,Landau_vol9}, the standard grand canonical ensemble (c.f. SI Appendix, Section 5A) gives a $T^d$ growth of the uncondensed eigen-particle density [and thus the normal-fluid density via \eqref{normal_fluid_density}],
\begin{align}
\frac{\rho_\mathrm{n}}{m}&=\frac{\tilde{N}_\mathrm{n}}{\mathcal{V}} = \frac{1}{\mathcal{V}} \sum_{\mathbf{k}\neq\mathbf{0}}n_\text{B}(\beta\omega_\mathbf{k})
= \int d\omega~n_\text{B}(\beta\omega)~g(\omega) \nonumber \\
&\propto \frac{1}{ v_\mathrm{v}^d} \int d\omega~ n_\text{B}(\beta\omega)~\omega^{d-1}
\propto \frac{1}{ (v_\mathrm{v}\beta)^d} = \left(\frac{T}{ v_\mathrm{v}k_\mathrm{B}}\right)^d,
\label{T3_Nn}
\end{align}
where $n_{\text{B}}(\beta \omega)=(e^{\beta\omega}-1)^{-1}$ is the Bose-Einstein distribution function at the inverse temperature $\beta=(k_{\text{B}}T)^{-1}$.
Here, the density of states, $g(\omega)\propto v_\mathrm{v}^{-d}\omega^{d-1}$, is used for a $d$-dimensional system with linear dispersion, $\omega_\mathbf{k}=v_\mathrm{v}|\mathbf{k}|$, in its eigen-particle-hole excitation spectrum.
(Note that at $T\to0$ the temperature dependence of the stiffness, $v_\mathrm{v}$, itself only leads to subdominant higher-order corrections to the depletion.)
Correspondingly, the superfluid fraction,
\begin{align}
\rho_\mathrm{s}(T)/\rho = [N-\tilde{N}_\mathrm{n}(T)]/N = 1 - AT^d
\label{T3_NnMM}
\end{align}
has a $T^d$-depletion (with coefficient $A$).

More realistically, in experiments used for Fig.~\ref{expfit}, particle fluctuations involving the thermal bath are actually negligible.
We therefore have also confirmed under the \textit{canonical} ensemble (c.f. SI Appendix, Section 5B) the same $T^d$ growth of the normal fluid fraction and \textit{independently} an identical depletion of the superfluid fraction as a generic property for 3D superfluidity.
}

\showmatmethods{} % Display the Materials and Methods section

\dataavail{
%To allow others to replicate and build on work published in PNAS, authors must make materials, data, and associated protocols, including code and scripts, available to readers in a public repository upon publication. Restrictions on full or partial access to these materials and requests for legal, ethical, and logistical (e.g., size) exceptions must be noted at submission. If requested, these materials must be made available to editors and reviewers during submission for the purpose of evaluating the manuscript. A statement detailing sharing plans will be included in the published article as provided within the submission form. Research datasets, whether original or previously published, must be cited in the references as a condition for publication. Please refer to our full \href{https://www.pnas.org/author-center/editorial-and-journal-policies\#materials-and-data-availability}{policy}.
Data used in this study are included in the main text and Supporting Information. All previously published data used in this study are cited in the reference list.
}

\acknow{During revision of this manuscript, in response to the preprint of this study, a recent study~\cite{Boris_prl,Boris_prb} discussed the need for a \textit{quantitative} renormalization of the linear-response superfluid density due to lattice effects (while still assuming an effective Galilean invariant \textit{unconstrained} field of the order parameter at the long space-time scales).
Nonetheless, the same universal $T^4$ thermal depletion of the linear-response superfluid density still results (naturally from the predetermined fixed point) and remains distinct from the hydrodynamic superfluidity focused on in our manuscript.
We thank P.B. Wiegmann, B. Svistunov, N. Prokof'ev, V. Grinenko, A. M. Garcia-Garcia, and Wei Wang for fruitful discussions on the traditional picture of superfluid density and the significance of recovering the proper hydrodynamic picture of the superflow.
This work is supported by the National Natural Science Foundation of China (NSFC) under Grant Nos. 12274287 and 12042507, and Innovation
Program for Quantum Science and Technology No. 2021ZD0301900.}

\showacknow{} % Display the acknowledgments section

\bibsplit[27]
%Use \bibsplit to split the references from the body of the text. Value "[3]" represents the number of reference in the left column (Note: Please avoid single column figures & tables on this page.)

% Bibliography
\bibliography{PNAS_Manuscript_references}

\end{document}